\documentclass[review]{elsarticle}

\usepackage{lineno,hyperref}

\usepackage{comment}
\usepackage{amsmath}
\usepackage{booktabs}
\usepackage{multirow}
\usepackage{graphicx}

\journal{Parallel Computing}

\begin{document}

\begin{frontmatter}

\title{Failure Analysis and Quantification for Contemporary and Future Supercomputers\tnoteref{mytitlenote}}
\tnotetext[mytitlenote]{This work was performed at the Ultrascale Systems Research Center (USRC) at Los Alamos National Laboratory (LANL), assigned the LANL identifier LA-UR-17-?????.}


\author{Li Tan\corref{mycorrespondingauthor}}
\cortext[mycorrespondingauthor]{Corresponding author}
\ead{darkwhite29@gmail.com}

\author{Nathan DeBardeleben}


\address{Ultrascale Systems Research Center, Los Alamos National Laboratory, Los Alamos, New Mexico, USA}

\begin{abstract}
Large-scale computing systems today are assembled by numerous computing units for massive computational capability needed to solve problems at scale, which enables failures common events in supercomputing scenarios. Considering the demanding resilience requirements of supercomputers today, we present a quantitative study on fine-grained failure modeling for contemporary and future large-scale computing systems. We integrate various types of failures from different system hierarchical levels and system components, and summarize the overall system failure rates formally. Given that nowadays system-wise failure rate needs to be capped under a threshold value for reliability and cost-efficiency purposes, we quantitatively discuss different scenarios of system resilience, and analyze the impacts of resilience to different error types on the variation of system failure rates, and the correlation of hierarchical failure rates. Moreover, we formalize and showcase the resilience efficiency of failure-bounded supercomputers today.
\end{abstract}


\end{frontmatter}


\section{Introduction}

Due to the demanding need of High Performance Computing (HPC) and the fast-advancing HPC technology, large-scale computing systems today are assembled by a large amount of computing units equipped with supporting components for an extremely computational and reliable HPC eco-system. Various studies showcase that failures are not rare events in such HPC systems due to the numerous interconnected components. Regardless of the fact that the growing number of components in HPC systems aggregate failure rates overall, root causes of failures in supercomputers include radiation-induced effects such as particle strikes from cosmic radiation, circuit aging related effects, and faults due to chip manufacturing defects and design bugs \cite{resilience_pattern}. Most failures remain undetected during post-silicon validation and eventually manifest themselves during the operation of HPC systems, e.g., in runs of HPC applications, and in upgrades or maintenance of system software and devices. As process technology continues to shrink and HPC systems today tend to operate at low supply voltage for power efficiency purposes, e.g., near-threshold voltage computing \cite{dac12}, hardware components of supercomputers become more susceptible to all types of faults at a greater rate. Therefore, Mean Time To Failure (MTTF) of the system is expected to dramatically decrease for forthcoming exascale supercomputers.

Resilience of HPC systems to various types of failures has become a first-class citizen in building scalable and cost-efficient HPC systems. In general, it is expensive to detect and correct such failures in large-scale computing systems in the presence of resilience techniques due to: (a) software costs, e.g., performance loss of the applications due to additional resilience code for calculating checksums/residues and saving checkpoints, and (b) hardware costs, e.g., extra components needed for modular redundancy like ECC memory and more disk space for checkpoint storage. Generally, resilience requires different extent of redundancy at various system levels in both time and space. Numerous studies have been conducted to improve the efficiency of existing resilience techniques for HPC systems. State-of-the-art solutions include Algorithm-Based Fault Tolerance (ABFT) \cite{ppopp12} and scalable multi-level checkpointing (SCR) \cite{sc10}. However, there exists lack of investigation to holistic failure analysis and fine-grained failure quantification for large-scale computing systems, covering realistic resilience scenarios in supercomputers up to date, which is definitely beneficial to understand failure pattern/layout of operational supercomputers today for better devising more effective and efficient fault tolerance solutions.

In this paper, we propose to discuss various types of errors and failures from different architectural levels of supercomputer architectures today, and quantify them into an integrated failure model to summarize overall system failure rates hierarchically, in different HPC scenarios. The primary contributions of this paper include: (a) study the quantitative correlation of failure rates among different components, different failure types, and different system layers of supercomputers, under specific overall system failure rate bounds, (b) discuss the quantitative impacts of system resilience levels (referred to as \emph{significance index} in the later text) to overall system failure rates, and (c) formalize the resilience efficiency of failure-bounded HPC systems.

The remainder of the paper is organized as follows: Section 2 introduces background knowledge. Section 3 discusses empirical failure models used in this work, and a holistic quantitative study (a refined failure model included) on failure-bounded supercomputers is presented in Section 4. Section 5 discusses related work, and Section 6 concludes.

\section{Background}

\begin{figure}[h]
\vspace{-1mm}
\centering
\includegraphics[width=2.79in]{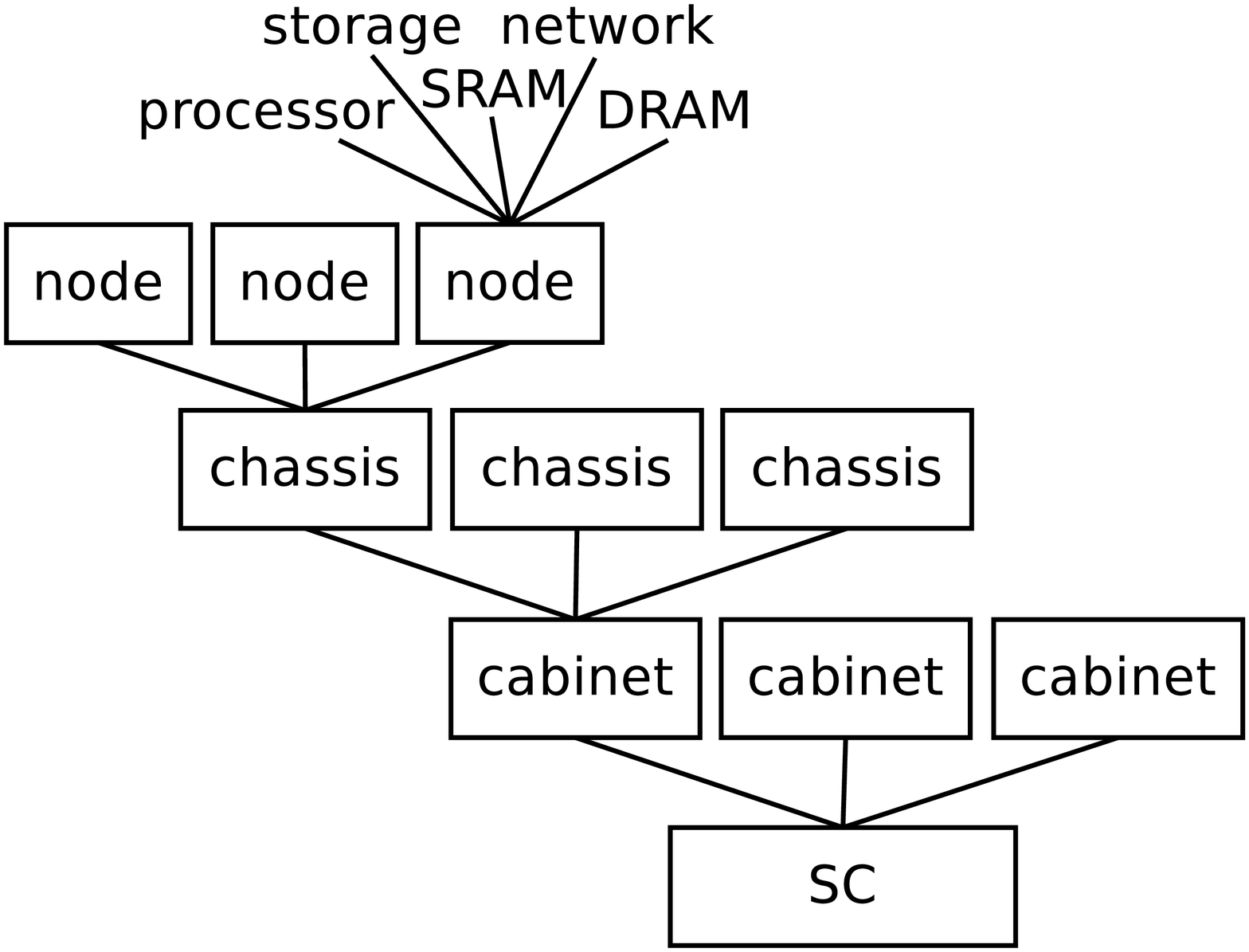}
\caption{Architectural Overview (Hierarchically) of Contemporary Supercomputers.}
\label{system_overview}
\vspace{-1mm}
\end{figure}

Supercomputers today is an extremely parallel and complex integration of numerous components, primarily categorized into computing units, network, storage, and supporting devices, e.g., cooling infrastructure, cables, and power supply. Figure \ref{system_overview} overviews the hardware architecture of contemporary supercomputers hierarchically (taking the supercomputer Trinity \cite{trinity} at Los Alamos National Laboratory for example, which ranks $10^{th}$ in the latest TOP500 list \cite{top500}): From top to down, a supercomputer is comprised of a number of cabinets (or racks), denoted as $N_{cabinet}$; each cabinet is comprised of a number of chassis (or blades), denoted as $N_{chassis}$; each chassis is comprised of a number of compute nodes, denoted as $N_{node}$ (without loss of generality, we ignore that there may exist a very small number of head nodes in the system that mostly do the management work). Finally, each computer node consists of hardware components including processors, storage, network, SRAM (on-chip), and DRAM (off-chip). According to the TOP500 list, top-ranked supercomputers up to date have hundreds of cabinets, thousands of chassis, and hundreds of thousands of compute nodes overall. In the figure, we only illustrate component details for one node (interconnects and other devices between nodes, chassis, and cabinets are omitted due to space limitation), and assume that all nodes and counterpart components (e.g., all cables) in the system are homogeneous (and thus have equivalent susceptibility to failures) to simplify our discussion.

Note that in Figure \ref{system_overview}, we use simplified terms to demonstrate the node configuration. Specifically, processors can be CPU and/or accelerators such as GPU and co-processors, which include functional units and control units. SRAM (on-chip) refers to registers and caches, and DRAM (off-chip) refers to main memory. Storage consists of any types of hard disk drives, solid-state drives, non-volatile memory, and cloud-based storage units. Network can be a high-speed interconnect such as InfiniBand.

\section{Failure Model}

Based on the system architecture shown in Figure \ref{system_overview}, we denote the failure rate of each level of system hierarchy as $\lambda_{sys}$, $\lambda_{cabinet}$, $\lambda_{chassis}$, and $\lambda_{node}$ individually.

\vspace{-2mm}
\begin{equation}
\label{system_failure_rate}
\lambda_{sys} = \lambda_{cabinet}N_{cabinet} = (\lambda_{chassis}N_{chassis})N_{cabinet} = (\lambda_{node}N_{node})N_{chassis}N_{cabinet}
\end{equation}
\vspace{-1mm}

As formulated in Equation (\ref{system_failure_rate}), it is straightforward to calculate the overall failure rate $\lambda_{sys}$ for an HPC system illustrated in Figure \ref{system_overview} based on the probability theory. It essentially shows failures are distributed over all available nodes in the system. We assume there are no idle nodes from each level of hierarchy when we consider failures, and thus all nodes are probabilistically equivalent for all types of errors.

\vspace{-2mm}
\begin{equation}
\label{node_failure_rate}
\lambda_{node} = \alpha\lambda_{soft} \oplus \beta\lambda_{hard}
\end{equation}
\vspace{-1mm}

For a compute node in supercomputers as shown above, there are two types of induced faults by nature: \emph{soft} errors and \emph{hard} errors. The former are \emph{transient} (e.g., memory bit-flips and logic circuit miscalculation), while the latter are usually \emph{permanent} (e.g., node crashes from dysfunctional hardware and system abort from power outage). We denote the failure rate of soft errors and hard errors as $\lambda_{soft}$ and $\lambda_{hard}$ respectively. In Equation (\ref{node_failure_rate}), we formulate the nodal failure rate as the integration ($\oplus$) of $\lambda_{soft}$ and $\lambda_{hard}$ (note that instead the mathematical addition ($+$) is not used here, given the different nature between soft errors and hard errors).

The parameters $\alpha$ and $\beta$ by $\lambda_{soft}$ and $\lambda_{hard}$ individually are referred to as the \emph{significance index} (SI) of failure rates. For various HPC systems equipped with different hardware and software resilient techniques, the SI of $\lambda_{soft}$ and $\lambda_{hard}$ varies. In general, SI represents the resilience to failures of a given system, and it has a negative correlation with failure coverage of the resilient techniques employed in the system, i.e., the more resilient the system is, the more failures can be recovered, the less SI value is. Consequently in Equation (\ref{node_failure_rate}) the nodal failure rate $\lambda_{node}$ changes accordingly, with the SI values introduced.

Due to the demanding requirements of system-wise power efficiency and resilience as the goal of US Department of Energy (DOE) for the upcoming exascale computers \cite{exascale}, current and future large-scale HPC systems needs to be not only power-bounded, but also failure-bounded, which means the overall system failure rate needs to be capped under a threshold value $\lambda_{sys}^{cap}$, provided a power budget \cite{powerbudget}. For simplicity of discussion, we define $\oplus$ by explicitly summing up soft and hard error rates. Therefore, based on Equations (\ref{system_failure_rate}) and (\ref{node_failure_rate}), we can reformulate the capped failure rates for soft errors and hard errors, under the specified expected system failure rate cap $\lambda_{sys}^{cap}$ below:

\vspace{-2mm}
\begin{equation}
\label{system_failure_rate_capped}
\lambda_{sys}^{cap} = (\alpha\lambda_{soft}^{cap} + \beta\lambda_{hard}^{cap})N_{node}N_{chassis}N_{cabinet}
\end{equation}
\vspace{-1mm}

According to the definition of soft errors and hard errors, node-wise we assume that processors, on-chip SRAM, and off-chip DRAM are the primary sources of soft errors, and storage and network are the main contributors to hard errors (in practice power supply contributes to hard errors considerably as well, which will be covered in the refined failure model in Section 4.3 where we assume power supply faults occur at chassis and cabinet levels). Without loss of generality (more components can be incorporated if needed), we look into the components above within a node, and formulate $\lambda_{soft}$ and $\lambda_{hard}$ more specifically as follows:

\vspace{-2mm}
\begin{equation}
\label{soft_error_rate}
\lambda_{soft} = \lambda_{processor} + \lambda_{SRAM} + \lambda_{DRAM}
\end{equation}

\vspace{-5mm}
\begin{equation}
\label{hard_error_rate}
\lambda_{hard} = \lambda_{storage} + \lambda_{network}
\end{equation}
\vspace{-1mm}


\section{Failure-bounded Quantitative Study}

In this section, we conduct exploratory quantitative discussion on several common scenarios in state-of-the-art HPC systems. With the established failure models above, our goals include: (a) given acquired failure data of system components, make some inferences on unknown failure rate caps of other components, and (b) speculate the system-/component-wise failure rate ranges under some known failure rate caps.

\subsection{Capping Failures by Types}

Per the mechanism of detection and correction, soft errors can be categorized as \emph{Detected and Corrected Errors} (DCE), \emph{Detected but Uncorrectable Errors} (DUE), and \emph{Silent Errors} (SE) \cite{ijhpca14}. Any unmasked SE are referred to as \emph{Silent Data Corruption} (SDC), i.e., incorrect program outputs. DCE generally occur in ECC-protected SRAM/DRAM, and examples of DUE include \emph{crashes} and \emph{hangs} of program execution. We assume that each compute node has statistically equivalent chances for incurring soft errors and/or hard errors. Moreover, assume an HPC system where on average 80\% of soft errors occurring in a single node are DCE (masked by ECC memory), 5\% are DUE, and 15\% are SDC, and there are no fault tolerance support at the software stack such as Algorithm-Based Fault Tolerance (ABFT), which means that 20\% of the total incurred soft errors circumvent resilience techniques employed, i.e., $\alpha$ = 0.2. Likewise, we assume that at the hardware stack, appropriate hardware-based resilience techniques are employed, and 60\% hard errors can be successfully masked, i.e., $\beta$ = 0.4.

Substituting $\alpha$ = 0.2 and $\beta$ = 0.4 into Equation (\ref{system_failure_rate_capped}) yields:

\vspace{-2mm}
\begin{equation}
\label{system_failure_rate_capped_eg1}
\lambda_{sys}^{cap} = (0.2\lambda_{soft}^{cap} + 0.4\lambda_{hard}^{cap})N_{node}N_{chassis}N_{cabinet}
\end{equation}
\vspace{-1mm}

Given that $N_{node}N_{chassis}N_{cabinet}$ is a constant number that refers to the total number of active compute nodes system-wide, and the assumed values for $\alpha$ and $\beta$, for the three remaining variables in Equation (\ref{system_failure_rate_capped_eg1}), we can easily solve one provided the other two.

The failure rate $\lambda$ can be expressed in terms of either Mean Time To Failure (MTTF) \cite{fgcs06} or Failure In Time (FIT) \cite{fpga05}. FIT is inversely proportional to MTTF and is defined as a failure rate of 1 in a billion hours. Here we adopt FIT as the calculation unit due to its additive nature, different from MTTF. Existing studies demonstrate that for HPC architectures nowadays, SRAM failure rates range from 10 FIT to 100 FIT \cite{dac11}, and DRAM failure rates are of the order of magnitude of 100 FIT \cite{isca13}. Therefore, without loss of generality, assume that there is a supercomputer of 100,000 nodes, with $\lambda_{soft}^{cap}$ = 200 FIT. Meanwhile, as a premise, $\lambda_{sys}^{cap}$ cannot exceed 5,000,000 FIT as required for system-level resilience. With the parameters already known, we can solve $\lambda_{hard}^{cap}$ as below:

\vspace{-2mm}
\begin{align}
\label{system_failure_rate_capped_eg2}
\lambda_{hard}^{cap} &= \frac{\frac{\lambda_{sys}^{cap}}{N_{node}N_{chassis}N_{cabinet}} - 0.2\lambda_{soft}^{cap}}{0.4}\nonumber\\
&= \frac{\frac{5000000}{100000} - 0.2 \times 200}{0.4} = 25
\end{align}
\vspace{-1mm}

\noindent which indicates that in order to achieve $\lambda_{sys}^{cap}$ no greater than 5,000,000 FIT, given $\lambda_{soft}^{cap}$ = 200 FIT and the above $\alpha$ and $\beta$ values, the threshold value of $\lambda_{hard}^{cap}$ is 25 FIT.

\begin{figure}[h]
\vspace{-4mm}
\centering
\includegraphics[width=3.79in]{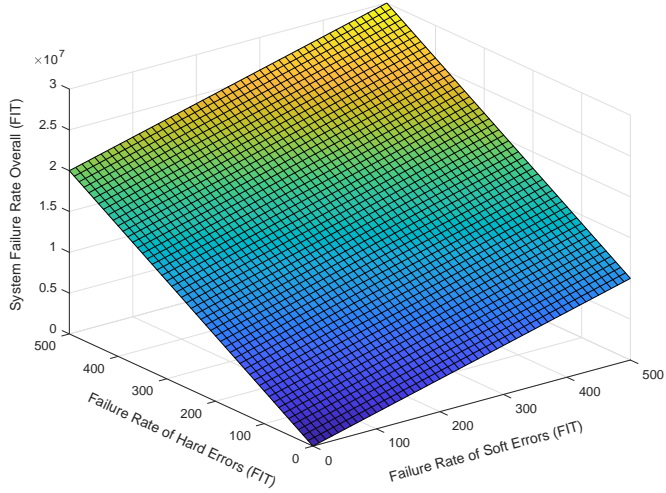}
\caption{Overall System Failure Rates of HPC Systems (Scenario 1).}
\label{system_failure_rate_capped_config1}
\vspace{-2mm}
\end{figure}

\vspace{1mm}
\noindent\textbf{\underline{Scenario 1}}: \textsf{An HPC System with Higher Resilience to Soft Errors}
\vspace{1mm}

Figure \ref{system_failure_rate_capped_config1} depicts the system failure rate curve, as nodal soft/hard error rate changes, provided the hypothesized failure rate SI values $\alpha$ = 0.2 and $\beta$ = 0.4 in Equation (\ref{system_failure_rate_capped}). This scenario represents HPC systems that have higher resilience to soft errors, compared to hard errors. We can see that although overall $\lambda_{sys}^{cap}$ is linear to $\lambda_{soft}^{cap}$ and $\lambda_{hard}^{cap}$ respectively, the system characteristic of higher resilience to soft errors makes $\lambda_{sys}^{cap}$ be affected more by the variation of $\lambda_{hard}^{cap}$, compared to that of $\lambda_{soft}^{cap}$. Figure \ref{system_failure_rate_capped_config1} also shows that this trend remains the same for all $\lambda_{soft}^{cap}$ and $\lambda_{hard}^{cap}$ values.

\begin{figure}[h]
\vspace{-4mm}
\centering
\includegraphics[width=3.79in]{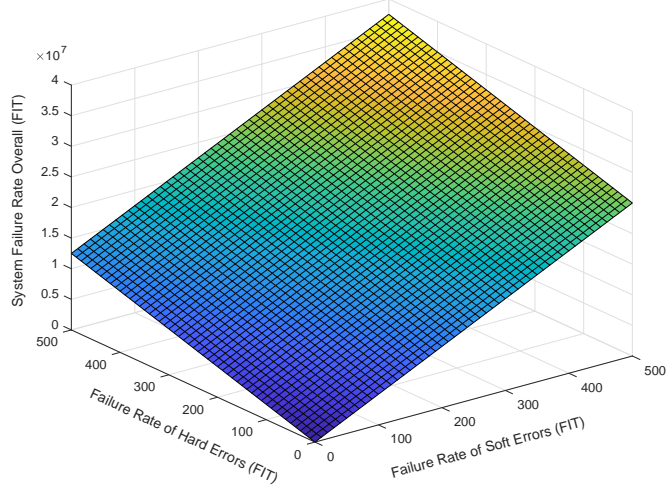}
\caption{Overall System Failure Rates of HPC Systems (Scenario 2).}
\label{system_failure_rate_capped_config2}
\vspace{-2mm}
\end{figure}

\vspace{1mm}
\noindent\textbf{\underline{Scenario 2}}: \textsf{An HPC System with Higher Resilience to Hard Errors}
\vspace{1mm}

Figure \ref{system_failure_rate_capped_config2} plots the system failure rate curve with another set of configuration of failure rate SI values $\alpha$ = 0.5 and $\beta$ = 0.25 in Equation (\ref{system_failure_rate_capped}), which reflects HPC system with higher resilience to hard errors instead of soft errors. Likewise, due to the higher tolerance to hard errors, the curve shows the trend that $\lambda_{sys}^{cap}$ tend to be impacted more by $\lambda_{soft}^{cap}$ instead of $\lambda_{hard}^{cap}$, i.e., with the same amount of change between $\lambda_{soft}^{cap}$ and $\lambda_{hard}^{cap}$, $\lambda_{sys}^{cap}$ varies greater with the change of $\lambda_{soft}^{cap}$, as shown in Figure \ref{system_failure_rate_capped_config2}.

\subsection{Capping Failures by Components}

Instead of capping failure rates of soft/hard errors at system level, HPC systems today also have resilience requirements for specific components. Given the system-wise failure cap and some acquired failure data from other components, we can obtain the capped failure rates for the interested components.

Substituting Equations (\ref{soft_error_rate}) and (\ref{hard_error_rate}) into Equation (\ref{system_failure_rate_capped}) yields:

\vspace{-2mm}
\begin{equation}
\label{system_failure_rate_capped_eg3}
\lambda_{sys}^{cap} \hspace{-0.4mm}=\hspace{-0.4mm} \left(\alpha(\lambda_{processor}^{cap} \hspace{-0.6mm}+\hspace{-0.6mm} \lambda_{SRAM}^{cap} \hspace{-0.6mm}+\hspace{-0.6mm} \lambda_{DRAM}^{cap}) \hspace{-0.6mm}+\hspace{-0.6mm} \beta(\lambda_{storage}^{cap} \hspace{-0.6mm}+\hspace{-0.6mm} \lambda_{network}^{cap})\right)\hspace{-0.6mm}N_{node}N_{chassis}N_{cabinet}
\end{equation}
\vspace{-1mm}

Likewise, $N_{node}N_{chassis}N_{cabinet}$ is a constant number. We employ the same hypothesized failure rate SI values as \textbf{Scenario 1}, $\alpha$ = 0.2 and $\beta$ = 0.4, and the same premise of an HPC system of 100,000 nodes with $\lambda_{sys}^{cap}$ = 5,000,000 FIT. In addition, we assume that from system logs historically, failure data of processor, SRAM, and network are acquired as follows: $\lambda_{processor}$ = 90 FIT, $\lambda_{SRAM}$ = 70 FIT, and $\lambda_{network}$ = 20 FIT. Substituting all known parameters into Equation (\ref{system_failure_rate_capped_eg3}), we have:

\vspace{-2mm}
\begin{align}
\label{system_failure_rate_capped_eg4}
0.2(90 + 70 + \lambda_{DRAM}^{cap}) + 0.4(\lambda_{storage}^{cap} + 20) &= 50\nonumber\\
32 + 0.2\lambda_{DRAM}^{cap} + 0.4\lambda_{storage}^{cap} + 8 &= 50\nonumber\\
\lambda_{DRAM}^{cap} + 2\lambda_{storage}^{cap} &= 50
\end{align}
\vspace{-1mm}

\noindent which indicates that in order to preserve the assumed failure rates, the quantitative relationship between $\lambda_{DRAM}^{cap}$ and $\lambda_{storage}^{cap}$ in (\ref{system_failure_rate_capped_eg4}) must be satisfied.

\subsection{Refining Failure Model from System Hierarchy}

Although an HPC system is comprised of compute nodes, failures may happen not only at local nodes, but also interconnects between nodes, power supply and other devices at chassis or cabinet level. When such failures occur at higher levels rather than at a single node, all nodes at the related levels are affected. For example, if the power supply at cabinet level fails, all nodes within the affected cabinets will be down. Consider the occurrence of failures hierarchically at different system layers, we refine the failure models as follows:

\vspace{-2mm}
\begin{equation}
\label{system_failure_rate_refined}
\lambda_{sys}^{ref} = \alpha'\lambda_{node}N_{node}^{total} + \beta'\lambda_{chassis}^{\overline{node}}N_{chassis}^{total} + \gamma'\lambda_{cabinet}^{\overline{node}}N_{cabinet}^{total}
\end{equation}
\vspace{-1mm}

\noindent note that in Equation (\ref{system_failure_rate_refined}) the parameters $\lambda_{chassis}^{\overline{node}}$ and $\lambda_{cabinet}^{\overline{node}}$ are failure rates of non-node devices/components at chassis and cabinet levels respectively, the parameters $\alpha'$, $\beta'$, and $\gamma'$ are the SI of node, chassis, and cabinet failure rates respectively, and the constants $N_{node}^{total}$, $N_{chassis}^{total}$, and $N_{cabinet}^{total}$ individually refer to the total number of nodes, chassis, and cabinets in the system overall. From previous models, we have:

\vspace{-2mm}
\begin{align}
N_{node}^{total} &= N_{node}N_{chassis}N_{cabinet}\nonumber\\
N_{chassis}^{total} &= N_{chassis}N_{cabinet}\nonumber\\
N_{cabinet}^{total} &= N_{cabinet}\nonumber
\end{align}
\vspace{-1mm}

Recent studies on DOE supercomputers indicate that failures at system-wide component level play a significant role in the resilience of the system. From analyzing one-year system logs of the supercomputer Mira at Argonne Leadership Computer Facility of Argonne National Laboratory, the frequency of fatal events based on different components and categories has been clearly identified. According to the statistics from this study, although soft errors (mostly memory errors) at node level are the most frequently occurred failure type, failures on system-wide components amount to at least 39.47\% of all observed failures, as listed in Table \ref{failure_rate_category_mira} \cite{alcf}. We can group all off-node failures in terms of $\lambda_{chassis}$ and $\lambda_{cabinet}$ given the specific location of failures. For simplicity of discussion, failures occurred between chassis and between cabinets are considered into $\lambda_{chassis}$ and $\lambda_{cabinet}$ respectively.

\begin{table}[h]
\vspace{2mm}
\scriptsize
\centering
\caption{Category of Failure Rates by Components of the Supercomputer Mira.}
\label{failure_rate_category_mira}
\begin{tabular}{|c|c|c|}
\hline
Component & Failure Rate & Failure Location\\
(system-wide) & (over one year) & (on/off node)\\
\hline
compute unit & \multirow{2}{*}{53.95\%} & \multirow{2}{*}{on node}\\
(soft error) & & \\
\hline
card & 14.47\% & off node\\
\hline
cable & 8.55\% & off node\\
\hline
link module & 6.58\% & off node\\
\hline
process/daemon & 5.26\% & off node\\
\hline
coolant monitor & 4.61\% & off node\\
\hline
other & 6.58\% & N/A\\
\hline
\end{tabular}
\normalsize
\end{table}

In order to study the relationship among the failure rates at node, chassis, and cabinet level, under a predefined system failure rate cap and with resilience techniques employed, we consider the following scenario:

\begin{figure}[h]
\vspace{-4mm}
\centering
\includegraphics[width=3.79in]{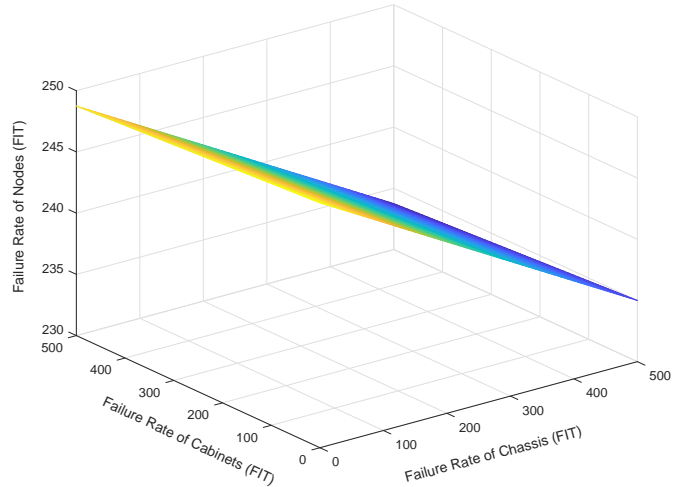}
\caption{Correlation of Node/Chassis/Cabinet Failure Rates under a System Failure Rate Cap (Scenario 3).}
\label{system_failure_rate_capped_config3}
\vspace{-2mm}
\end{figure}

\vspace{1mm}
\noindent\textbf{\underline{Scenario 3}}: \textsf{An HPC System with Resilience and Capped System Failure Rates}
\vspace{1mm}

Figure \ref{system_failure_rate_capped_config3} shows the node, chassis, and cabinet failure rate curve for another HPC system scenario, where we assume that the node, chassis, and cabinet failures in this system are tolerated to some extent by employed resilience techniques individually, and consequently $\alpha'$ = 0.2, $\beta'$ = 0.6, and $\gamma'$ = 0.5. We adopt the same system architectural configuration as previous examples: 100,000 nodes (100 nodes per chassis, 10 chassis per cabinet, and 100 cabinets in the system), with $\lambda_{sys}^{cap}$ = 5,000,000 FIT. Therefore, Equation (\ref{system_failure_rate_refined}) is instantiated below:

\vspace{-2mm}
\begin{align}
\label{system_failure_rate_refined_inst}
5000000 &= 0.2\lambda_{node} \times 100000 + 0.6\lambda_{chassis}^{\overline{node}} \times 1000 + 0.5\lambda_{cabinet}^{\overline{node}} \times 100\nonumber\\
100000 &= 400\lambda_{node} + 12\lambda_{chassis}^{\overline{node}} + \lambda_{cabinet}^{\overline{node}}
\end{align}
\vspace{-1mm}

Specifically, Figure \ref{system_failure_rate_capped_config3} is an illustrated version of Equation (\ref{system_failure_rate_refined_inst}). We can see that as $\lambda_{chassis}^{\overline{node}}$ and $\lambda_{cabinet}^{\overline{node}}$ change, the variation of $\lambda_{node}$ is comparatively small, i.e., $\lambda_{chassis}^{\overline{node}}$ and $\lambda_{cabinet}^{\overline{node}}$ both range from 0 to 500 FIT, while $\lambda_{node}$ ranges only from 230 to 250 FIT. This is because there exist much more nodes compared to chassis and cabinets in the system overall. However, statistically, failure rates of a single node are smaller than failure rates of a single chassis or a single cabinet. In general, with a capped system failure rate, the growing of failure rates of any hierarchy level (node, chassis, or cabinet), leads to the decreasing of failure rates of the other two levels. We can also see that the variation of $\lambda_{chassis}^{\overline{node}}$ has a greater impact on the variation of $\lambda_{node}$, compared to the variation of $\lambda_{cabinet}^{\overline{node}}$.

\subsection{Failure-bounded HPC System Time Usage and Resilience Efficiency}

\begin{figure}[h]
\vspace{-1mm}
\centering
\includegraphics[width=4.79in]{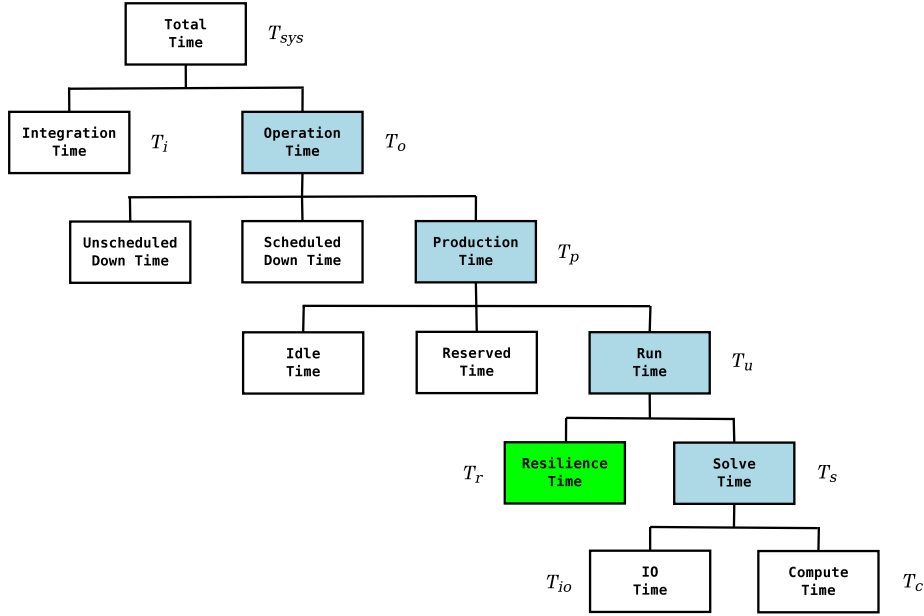}
\caption{Time Usage Overview of Failure-bounded HPC Systems.}
\label{system_time_dist}
\vspace{-1mm}
\end{figure}

Regarding the impacts of resilience on HPC systems, the breakdown of system time usage by functionality (e.g., system in idle, operation, computation, or I/O) is highly beneficial since fine-grained efficiency analysis is feasible. Figure \ref{system_time_dist} overviews the general time usage of typical HPC systems today \cite{ras}. We can clearly see that the time used for resilience purposes $T_r$ is a part of the system run time $T_u$, while the other part is solve time $T_s$ which is in general application-specific. Without loss of generality, we assume that in Figure \ref{system_time_dist} the highlighted time components ($T_o$, $T_p$, $T_u$, $T_r$, and $T_s$) account for the majority of the total system time $T_{sys}$. Furthermore, the resilience efficiency of an HPC system can be formalized as follows:

\vspace{-2mm}
\begin{equation}
\label{resilience_efficiency}
E_{res} = E_{res}^{run} \times E_{run}^{prod} \times E_{prod}^{oper} = \frac{T_r}{T_u} \times \frac{T_u}{T_p} \times \frac{T_p}{T_o} = \frac{T_r}{T_o}
\end{equation}
\vspace{-1mm}

Note that in practice, $T_u$ varies depending on if there exist failures or not in HPC runs. Since if there are no failures during HPC runs, no extra costs on recovering from failures which makes $T_u$ smaller. Specifically, let the system employ Checkpoint/Restart (C/R) as the resilience technique. If no hard errors occur while applications are running, the system does not need to restart from the last saved checkpoint, and then less time spent on resilience, i.e., smaller $T_u$ while $T_s$ unchanged. For example, assume that there is an HPC system in operation of 10,000 hours, where 8,000 hours in applications running without failures, while 8,400 hours in application running with failures. Without resilience techniques employed, the application total run time is 6,600 hours (which can also be estimated using application algorithmic complexity and computation capability of the system). Using Equation (\ref{resilience_efficiency}), we can easily obtain the resilience efficiency of both scenarios below:

\vspace{-2mm}
\begin{equation}
\label{resilience_efficiency_eg1}
E_{res}^{\overline{err}} = \frac{8000 - 6600}{10000} = 14\%
\end{equation}
\vspace{-1mm}

\vspace{-2mm}
\begin{equation}
\label{resilience_efficiency_eg2}
E_{res}^{err} = \frac{8400 - 6600}{10000} = 18\%
\end{equation}
\vspace{-1mm}

As shown, $E_{res}^{err}$ has a greater value than $E_{res}^{\overline{err}}$, due to the presence of failures which needs additional resilience time on recovering for correct HPC runs. For different resilience techniques, the difference between $E_{res}^{err}$ and $E_{res}^{\overline{err}}$ may vary, because of the different nature of recovering from failures.

It is well-studied that supercomputers today (up to petascale) are exposed to high failure rates due to various root causes, with MTTF ranging from 50 minutes to 230 minutes \cite{exascale_failure}. Forthcoming exascale supercomputers are expected to suffer from increased failure rates due to a greater number of components, with predicted MTTF ranging from 22 minutes to 120 minutes \cite{exascale}. With the expected failure rates, we can speculate the resilience efficiency of future exascale supercomputers using our models. Assume that there is an exascale system in operation of 10,000 hours, and one failure occurs every 120 minutes, with 40\% hard errors and 60\% soft errors. The employed resilience techniques can successfully capture every failure and take 0.7 hour and 0.2 hour to detect and recover from hard errors and soft errors individually. Using Equation (\ref{resilience_efficiency}), we calculate the resilience efficiency below:

\vspace{-2mm}
\begin{equation}
\label{resilience_efficiency_eg3}
E_{res} = \frac{0.7 \times (0.4 \times \frac{10000}{2}) + 0.2 \times (0.6 \times \frac{10000}{2})}{10000} = \frac{1400 + 600}{10000}  = 20\%
\end{equation}
\vspace{-1mm}

From the calculation shown above, we can see that in order to obtain higher resilience efficiency for failure-bounded HPC systems in this era, we need to develop more cost-effective resilience techniques, or increase the MTTF of future supercomputers.

\section{Related Work}

Modeling methods have been extensively used for large-scale computing systems, for the purposes of failure prediction \cite{ipdps12b} \cite{sc12}, trade-off optimization \cite{acsd14} \cite{taco16}, and vulnerability reduction \cite{ics12} \cite{dsn-w17}. Gainaru \textit{et al.} \cite{ipdps12b} proposed to characterizing the normal and faulty behavior of HPC systems by using signal analysis to model the flow of each state event during HPC system lifetime. The extracted models accurately reflected system outputs and improved the effectiveness of fault prediction. The subsequent work \cite{sc12} leveraged data mining techniques to offer an adaptive failure prediction module for accurate fault prediction, and was evaluated on two large-scale systems for prediction precision and recall impacts. Instead of focusing on analyzing the system state data (referred to as system events in \cite{ipdps12b} and \cite{sc12}), our work investigates failure rate correlation at different system hierarchical levels and system components levels. Rafiev \textit{et al.} \cite{acsd14} studied the interplay between critical dimensions in HPC, i.e., performance, energy, and reliability using a modeling framework based on a resource-driven graph representation. The layer-agnostic models applied efficiently to large-scale systems and diverse types of concurrency. Tan \textit{et al.} \cite{taco16} quantitatively modeled the integrated energy efficiency in terms of performance per Watt and showcased the trade-offs among typical HPC parameters, by extending the Amdahl’s Law and the Karp-Flatt Metric. The proposed models were evaluated to help find the optimal HPC configuration for the highest integrated energy efficiency with resilience. This work focuses on the resilience of HPC systems only and our failure model is based on the probability theory. Casas \textit{et al.} \cite{ics12} presented an approach that analyzes the vulnerability of sparse scientific applications to hardware faults at large scales, and reduced their vulnerability by protecting the most vulnerable components and failure prediction. Leveraging register vulnerability, Tan \textit{et al.} \cite{dsn-w17} investigated the validity of failure rates in HPC systems at near-threshold voltage, and empirically evaluated the power saving opportunities without incurring observable number of soft errors during HPC runs. Our work differs from them since the proposed model here is for better understanding failure pattern of operational supercomputer architectures today and thus devising more feasible resilience solutions accordingly.

\section{Conclusions}

Due to the expansion of HPC systems in size and duration in use, it is critical to maintain the resilience of supercomputers today. For resilience purposes, it is beneficial to quantify failures in existing failure-bounded HPC systems in a fine-grained fashion. In this paper, we conduct an exploratory quantitative study on holistic failure modeling for contemporary large-scale computing systems, which also sheds light on understanding potential failures on forthcoming supercomputers in the exascale era, and helps better devise more feasible resilience solutions at scale. Specifically, we integrate different failures from the perspective of system hierarchy, and summarize the overall system failure rate formally. We also discuss various scenarios of HPC system resilience categorized by error types, system components, and hierarchical levels, and formalize the significance index of failure rates and the resilience efficiency of supercomputers today under a system failure rate cap.

\bibliography{elsarticle-template}

\end{document}